# A Neutrino Beacon

A. A. Jackson*

*TRITON SYSTEMS, LLC, 17000 El Camino Real, Houston, TX 77058*

*"So the first rule of my game is: think of the biggest possible artificial activities with limits set only by the laws of physics and look for those."*

*— Freeman Dyson [1]*

**Abstract:**
Observational SETI has concentrated on using electromagnetism as the carrier, namely radio waves and laser radiation. Michael Hippke [2] has pointed out that it may be possible to use neutrinos or gravitational waves as signals. Gravitational waves demand the command of the generation of very large amounts of energy, Jackson and Benford [3]. This paper describes a beacon that uses beamed neutrinos as the signal. Neutrinos, like gravitational waves, have the advantage of extremely low extinction in the interstellar medium. To make use of neutrinos an advanced civilization can use a gravitational lens as an amplifier. The lens can be a neutron star or a black hole. Using wave optics one can calculate the advantage of gravitational lensing for amplification of a beam and along the *focal axis* it is exceptionally large. Even though the amplification is very large the diameter of the beam is quite small, less that a centimeter. This implies that a large constellation of neutrino transmitters would have to enclose the local neutron star or black hole to make an approximate isotropic radiator. The operational energy needed is about .01 Solar, this means that such a beacon would have to be built by a Kardashev Type II civilization.

## 1. Introduction

The Search for Extra-Terrestrial Intelligence has been going on now for over 50 years. These searches have focused on using photons in the infra-red, microwave and radio-frequencies. These ideas spawned concepts that advanced civilizations might build beacons. Recently thoughts about the spectrum of long range carriers have been extended to neutrinos and gravitational waves, Hippke, 2018.

| Carrier | Rest Mass Mev/c² Speed | Lifetime | Extinction | Sources |
|---|---|---|---|---|
| Photons | 0<br><br>v = 1 | Stable | .001 | Beacons<br>Deliberate Transmissions<br>Propulsion Stations<br>Ambient Radiation from Star Ships<br>Waste Heat Star Ships<br>Waste Heat ("Dyson Spheres")<br>A large "Instrumentality" |
| neutrinos | ~.001<br><br>v~1 | Oscillations<br><br>Stable | ~0 | From Star Ships<br>A Beacon<br>Deliberate Beaming for Communication |
| Gravitons Gravitational Waves | 0<br><br>v=1 | Stable | ~0 | A very advanced civilization's beacon |

Table 1  Long range zero and near zero rest mass messengers (after Hippke[2])

Gravitational waves as a signal beacon has been presented by Jackson and Benford [3]. Over cosmic distances directed gravitational waves are very expensive in energy, demanding a command of levels of energies above Kardashev III civilizations [4].

Pasachoff and Kutner[5], Learned, et al.[6] and Pakvasa [7] have presented ideas about using neutrinos for interstellar communication. Advantages of using neutrinos are: (1) neutrinos arrive almost without attenuation from any source direction, this would have big advantage in the Galactic plane, (2) Neutrinos, at the Earth, are rare in certain energy ranges, and from a given direction are all but negligible, (3) Even when photons are not completely blocked, their scattering introduces jitter in arrival time as well as direction [6].

Herein is presented a method for using a gravitational lens to amplify the effectiveness of a neutrino beam transmitted across interstellar distances.

## 2. Gravitational Lensing

Gravitational lensing has been proposed as a means of enhanced interstellar communication, see, Eshleman (1979)[8], or Maccone for an extensive treatment [9]. Consider a K2 civilization using a Schwarzschild or Kerr black hole as a means of focusing radiation from a beaming station. The advantage of this is the enormous amount of amplification possible. Any gravitating body may serve as a lens, an ordinary star, a neutron star or a black hole [10]. Note, under consideration here is using a gravitational lens as a *transmitter* not as a telescope.

It is known that focusing of light by a gravitating point mass leads to caustic focal structures and that a geometrical optics treatment gives an infinite gain of the focused radiation on the optical axis [11]. The optical axis is the line connecting the source, lens and observer; we are interested in the case where the source is behind the lens and there is an observer at the caustic crossing. In a wave optics treatment the gain on the caustic is not infinite but very large as was shown by Bliokh and Minakov (1975) [12]. We want to emphasize the observable amplification at the caustic crossing, thus some elaboration of the calculation is in order , below is a sketch of the derivation.

Gain is defined as the ratio of the scattered flux per unit solid angle (at a large distance from the scattering center) to incident radiation flux. The proper calculation of the gain on the optical axis must take diffraction into account. This can be done in two ways solving the wave equation as a scattering problem for the scalar case with the proper boundary conditions [13] or using the Fresnel–Kirchhoff phase integral [11]. Take the case of solving the wave equation. The space-time of a black hole to be described by the Schwarzschild metric [11],

$$ds^2 = g_{\mu\nu}dx^\mu dx^\nu = (1-\frac{r_s}{r})dt^2 - (1-\frac{r_s}{r})^{-1}dr^2 - r^2(d\theta^2 + \sin^2 d\varphi^2). \quad (1)$$

where r, θ and φ are spherical coordinates and t is the coordinate time, with M the mass of the black hole if G and c are gravitational constant and c the speed of light then

$$r_s = \frac{2GM}{c^2} \quad (2)$$

is the Schwarzschild radius.

Write the propagation of a plane wave on the Schwarzschild background and use the Fourier decomposition for the angular parts [11]

$$\varphi(x^\mu) = \frac{1}{4\pi r}\sum_l\sum_m e^{-i\omega t}Y_l^m(\theta,\phi)\Psi_{lm}(\omega,r), \quad (3)$$

where $Y^m_n$ are the spherical harmonics and Ψ solves the radial wave equation of a scalar field

$$\nabla^2\Psi + (\omega^2 + \frac{2\omega^2 r_s}{r})\Psi = 0, \quad (4)$$

that is, the wave equation of scalar radiation of frequency ω and wave function Ψ.
If (4) is subject to boundary conditions that define the incoming and scattered wave

$$\Psi = e^{ikz}(in) + f(\theta)\frac{e^{ikr}}{r}(scattered), \quad (5)$$

Then the solution, in cylindrical coordinates, is [12, 13, 14, 15, 16, 17, and 18]:

$$\Psi = e^{\pi\omega\frac{r_s}{2}}\Gamma(1-i\omega\frac{r_s}{2})\,_1F_1(i\omega\frac{r_s}{2},1;i\omega(r-z))e^{i\omega z}. \quad (6)$$

Where $_1F_1$ is a confluent hypergeometric function and r, θ and z are cylindrical coordinates.

In the high frequency limit ω→∞ and first order gravitational field, the wave function (2) and the incoming intensity $I_0$ define the gain [18],

$$gain = \frac{\Psi\Psi^*}{I_0} \quad (7),$$

And Ψ in this limit (6), close to the focal line in the image plane leads to [18]

$$gain_0 = \frac{4\pi^2 r_s}{\lambda} J_0^2(2\pi \frac{\rho}{\lambda}\sqrt{\frac{2r_s}{z}}) \cdot (8)$$

where $J_0$ is the first order Bessel function. The first term is the scaling, proportional to the Schwarzschild radius divided by the wavelength. The Bessel function term is oscillating and due to interference averages to ½. The second term also defines the spatial extent to a 'focal beam' (a caustic on the focal line this term is sometimes called the Point Spread Function, PSA. A very exhaustive treatment of this lensing problem for the wave optics electromagnetic case is given by Turyshev and Toth 2017, [18].

In this wave optics solution it was noted that there are three regions a shadow, a region of geometric optics and on the focal line a beam of large intensity due to diffraction, Figure 1 . From (4) plots can be made of the PSA as a function of ρ and z. There is a prime focal point at $z_0 = R_n^2/2r_s$ , and z down the axis can be written as

$$z(b) = z_0(b^2/R_n^2) \quad (9)$$

where b is the impact parameter of a neutrino ray and $R_n$ is the neutron star radius.
 Fix z at $z_0$ then in the image plane perpendicular to this point in normalized units the spread is given in figure 1

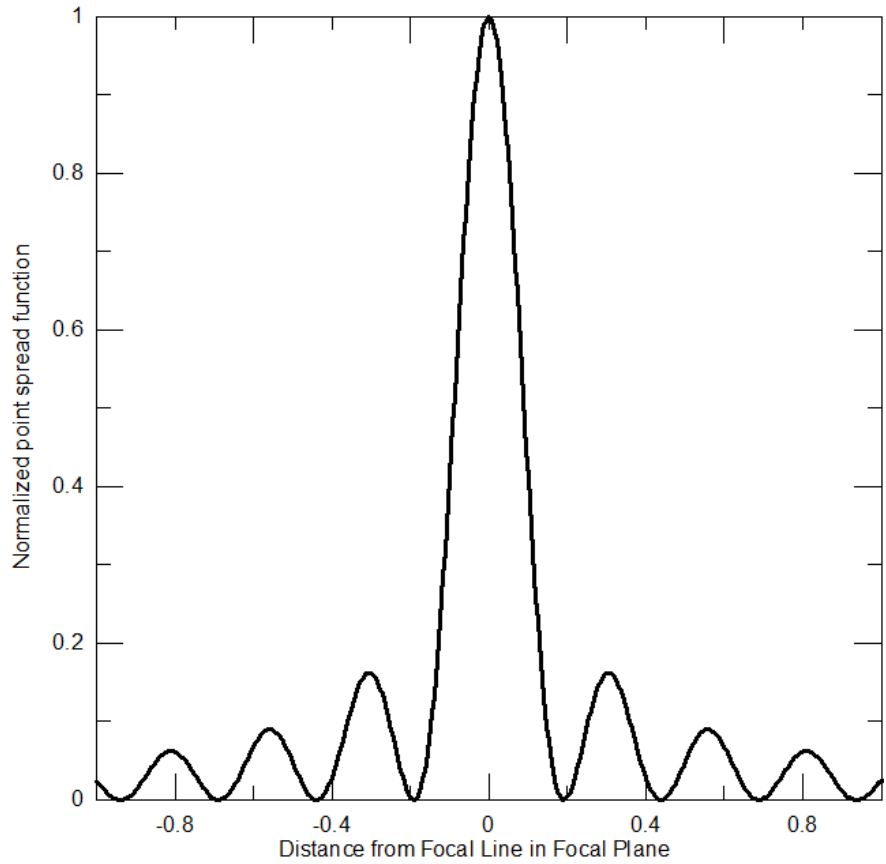

Figure1. The normalized point spread, PSF, for a gravitational lens.

If ρ is fixed and z allowed to increase down the focal axis the PSF becomes asymptotic in a few multiples of the prime focal point , see figure 2.

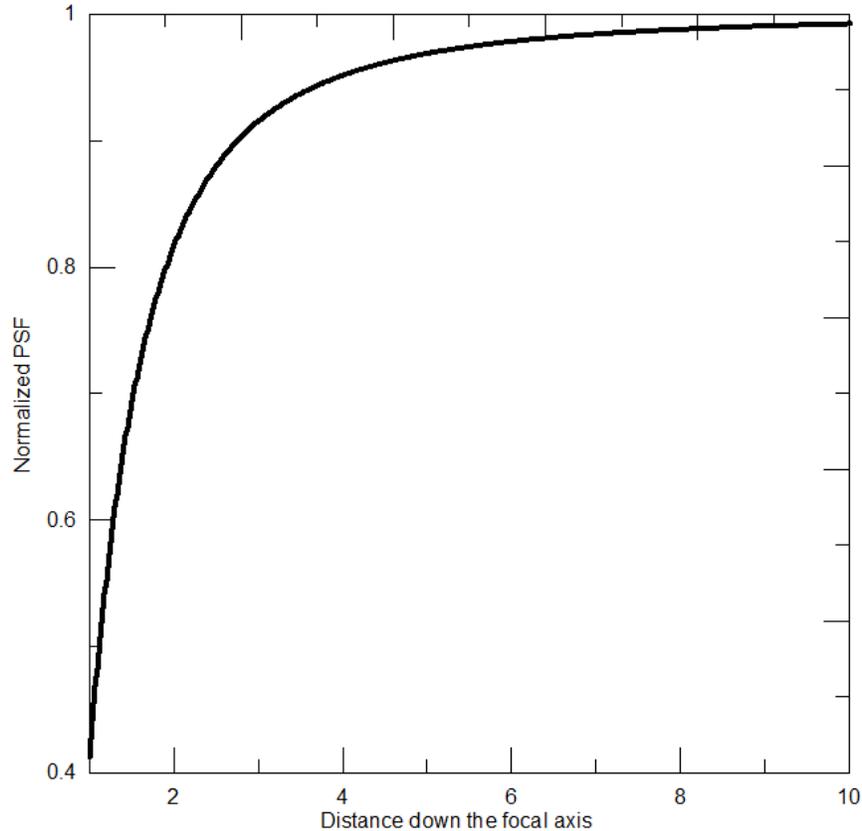

Figure 2. The PSF as a function of distance down the focal axis for a fixed $\rho$.

Then this implies that the central peak in figure 1 broadens and most of the flux falls in a region defined by the zeros of the Bessel function.
For a stellar mass gravitational lens and 1 Gev neutrinos, the wavelength is about $10^{-14}$ cm, the gain is approximately $10^{20}$ ! The characteristic radius of main region of concertation is about one micron; however there is an effective flux out to about one centimeter. This beam intensity extends to very long distances only diminished by absorption in the interstellar medium, encounters with a massive object like a planet or star and a very small beam divergence.
    The usual treatment above diffraction model is for a telescope, of interest here is a transmitter, that is forward scattering is used to propagate a beam of high concentration.

### 3.0 Transmission and Detection

    How might an advanced civilization configure an artifact to exploit the model described in section 2? The gravitation lens could be a neutron

star (a neutron star is very spherical) or a black hole (a non rotating black hole is spherical, or it could be an ordinary star however other condensed stellar objects are not considered here. Consider a neutron star with a neutrino beam transmitting station in orbit about it. At present a typical high-energy neutrino beam is made from the decay of π mesons. Large accelerators boost, usually protons, to relativistic energies which strike a target that produces pions and kaons which decay in flight into neutrinos, electrons and muons. The useful feature is that the pions and kaons can be focused and a directed beam of neutrinos produced. These neutrinos will emerge from the source into beam angle θ determined by Special Relativity, that is

$$\theta = \frac{1}{\gamma}. \quad (9)$$

Where γ is the Lorentz factor $\frac{1}{\sqrt{1-\beta^2}}$ (β=speed as a fraction of light speed). Taking the gravitational lens to be a neutron star or a non-rotating stellar mass black hole a transmitter is envisioned (a rotating black hole ,a Kerr black hole, could be used but the modeling is more complicated and is not considered here). Place a neutrino beam transmitter at 100 neutron star radii or at about 1000 km and using an impact parameter of 25 radii the opening angle (by simple geometery) will be approximately 1 degree, this gives a Lorentz factor of 74 which results in a pion bunch with energy of about 8 Gev. Thus there is a beam of approximately 8 Gev neutrinos incident on an annulus with a width of 15 km. Suppose the energy input is one watt for one second, then $8 \times 10^{14}$ neutrinos flow into an area of about $8 \times 10^{11}$ m² which in turn is focused into an area of approximately 1 cm² resulting in a bolt of Gev neutrinos. A 1.4 solar mass neutron star has a Schwarzschild radius of about 3 kilometers and 8 Gev neutrinos have a wavelength of about $10^{-14}$ cm ($10^{-4}$ picometres) thus the gain from (8) (or amplification) from the gravitational lens is about $10^{21}$ on the focal axis. On the focal axis in the center of the 'bolt' is a one cm spot with approximately $10^{20}$ neutrinos per pulse.

    Define the resolution of the neutron star beacon by the location of the first null of the Bessel function argument in (8) or the first zero in figure 1, then the 'spot' in the image plane is associated with the angle $\theta_s = \rho/z$, where ρ is the 'radius' of a spot at a transmitter distance of z. To emphasize

take as the region of interest the distance to the first null of the Bessel function, equation (8),

$$J_0^2(2\pi \frac{\rho}{\lambda}\sqrt{\frac{2r_s}{z}}) = 0 \quad (10)$$

The first null is $r_n = 2.40483$, then (10) can be solved for $\theta_s$ [17],

$$\theta_s = \frac{r_n \lambda}{2\pi b} \quad . \quad (11)$$

The radius of the spot at the detection distance is $\rho = \theta_s z_t$, where $z_t$ is the distance from transmitter to the target. Here is a problem similar to the 'resolution' of an image by a gravitational lens except the detector is a neutrino telescope.

The dependence in (11) is only on two independent parameters, wavelength and impact parameter, b. With the transmitter at 100 neutron star radii and the maximum impact parameter set at 25 neutron star radii the resulting neutrino energy ,(9), means with a wavelength of $10^{-14}$ cm, resulting in a very small angle, $\theta_s = 2.4 \times 10^{-22}$ radians.

As an example take the characteristic distance of an advanced civilization in our galaxy to be 10,000 light years then $\rho$ is approximately 2 centimeters!

If the transmitter pulses at a rate of one second then the 'bolt' of neutrinos moves off to infinity at almost the speed of light and only attenuated when it strikes a mass. Suppose it encounters a neutrino 'detector', estimate the number of scattering events in a cylinder of water 1 meter long and 1 meter in diameter, the number of events is

$$N_{events} = \frac{\sigma N_{in} \rho_w N_{av}}{A} \quad (12)$$

Where $\sigma$ is the interaction cross section for 8 Gev neutrinos , which is approximately $10^{-42}$ m² ($10^{-38}$ cm²) [19] , $N_{in}$ is the number of neutrinos passing in per unit time, $\rho_w$ is the density of water, $N_{av}$ is Avogadro's number and A is the atomic number of water. Even though the cross section is extremely small approximately an average of 2 events are seen per second in cylinder about 1 millimeter in diameter, figure 3.

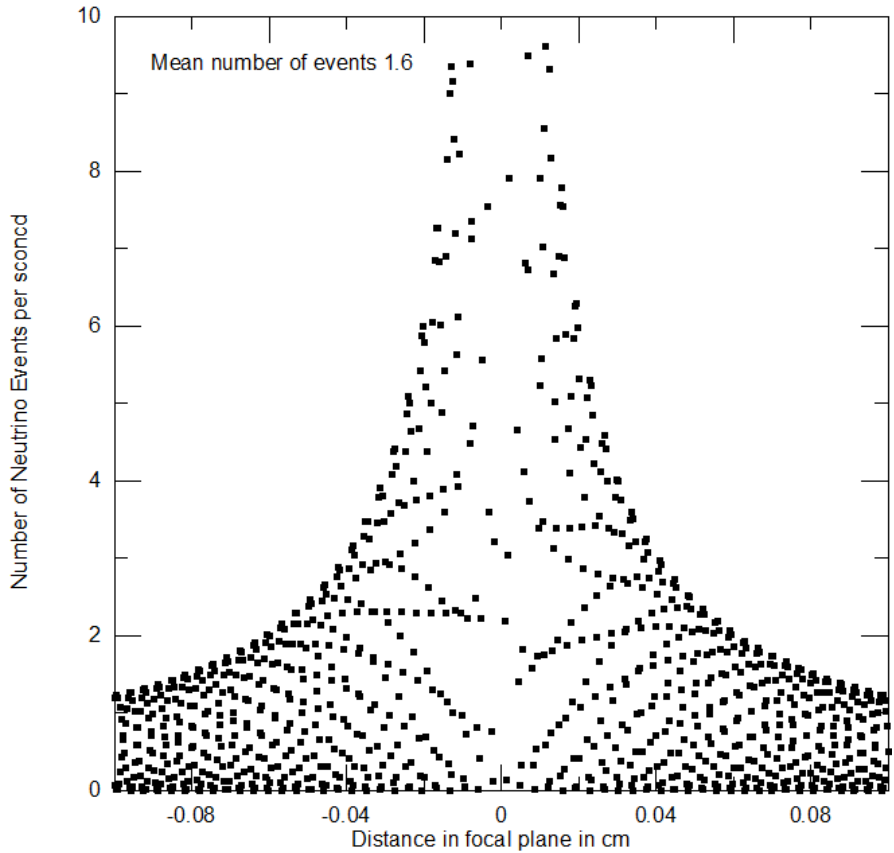

Figure 3. Neutrino events in a candidate neutrino water detector.

The configuration here is extremely idealized and there are a number of parameters subject to errors, but this is a lot of detections, even if it was one detection per hour. IceCube [20] detects about 11 neutrinos a day.

**4.0 The Beacon**

An extreme problem presents itself; a one centimeter beam aimed over 10,000 light has an almost zero probability of intersecting a detector on the Earth. A simple estimate of the probably of a detection is $\theta_s/4\pi$, in this case approximately $10^{-21}$. One could extrapolate that a Kardashev 2 (K2) civilization can construct the pointing accuracy of an ultra-advanced mechanism to aim an almost infinitively sharp beam at a target over thousands of light years with accuracy, but that will be left as a thought.

One can make another extrapolation, if the number of transmitters is increased to make the transmitter more isotropic-like it could make detection easier. Suppose that a K2 type civilization capable of interstellar flight can reach a neutron star or black hole , it should have the technological capability to build a beacon consisting of an array of transmitters in a constellation of orbits about the neutron star. Let this constellation consist of $10^{18}$ 'neutrino' transmitters 1 meter in characteristic size 'covering' the area of a sphere 1000 km in radius with $10^{18}$ particle accelerators in orbit figure 5. At the present time there is the development of plasma Wakefield particle accelerators that are meters in size [21, 22]. It is probable that a K2 civilization may construct Wakefield electron accelerators of very small size.

We will also suppose that a K2 civilization can solve all the problems of pointing errors and jitter, which could be substantial, the engineering physics is orders of magnitude beyond any our current civilization is capable of. Also the engineering physics involved in guidance, navigation and control of a constellation of $10^{18}$ small space craft will be extremely complicated.

To summarize, the engineering physics approach would be to build a constellation of neutrino beam transmitters. Place, in orbit, at 100 neutron star radii, $10^{18}$ advanced small Wakefield accelerators one meter long and 20 centimeters in diameter, figure 5., each point on figure 5 is occupied by an accelerator neutrino source, figure 6. Plasma-based accelerators are already producing high energy particle beams, what a K2 civilization may be capable of , for accelerators, is an extrapolation. With $10^{18}$ accelerators pointing four pi radians the probability of detection increases to approximately $10^{-3}$ and the detection rate at 10,000 light years becomes approximately 5 per minute. The power required for the whole artifact' is about .01 Solar, which is a K2 command of energy.

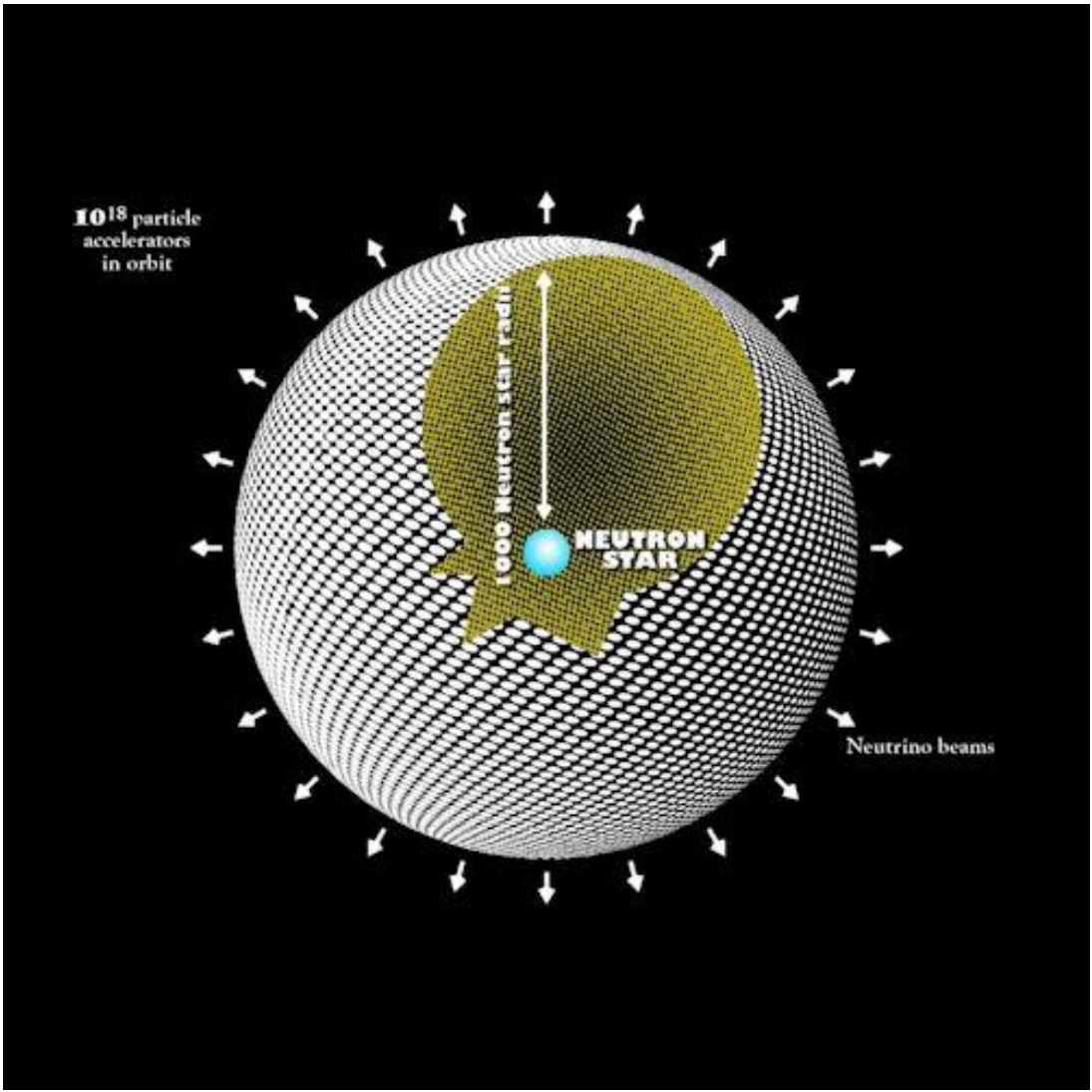

Figure 5: A representational constellation of $10^{18}$ accelerator-transmitters in orbit(nothing to scale).

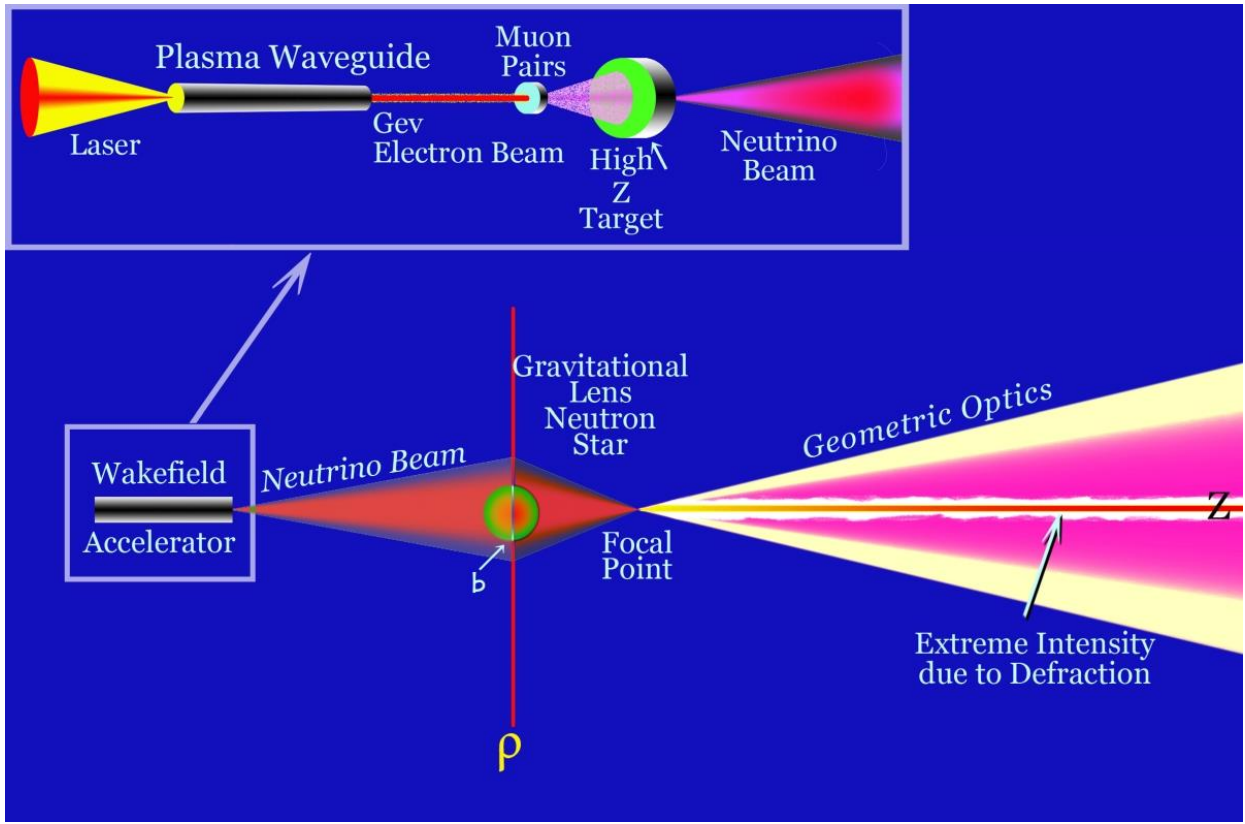

Figure 6: A schematic illustration of a possible neutrino accelerator-transmitter, the accelerator and lens (nothing to scale).

Of note, there is one constraint the advanced civilization must master. The transmitters in the orbital constellation must be configured as to appear to the gravitational lens as *point-sources*. It will be supposed that a K2 technology can master this technology.

It is also possible for the gravitational lens to be a black hole, however the results are nearly same except for 'glory' scattering, radiation scattered 'backward', which may be of advantage [23]. Scattering by a rotating black hole is very complicated matter and is not dealt with here.

## 5.0 Conclusions

An instrumentality has been presented for directed neutrino signal transmission. The 'beam' could encode information in neutrino beam by on-off keying, or more sophisticated modulation is possible [24]. An advanced civilization may deploy such a beacon as a 'honey-pot' attracting attention to an electromagnetic transmitter broadcasting more information in sophisticated manner. The 'artifact' presented here is a thought experiment; a Kardashev 2 civilization would likely have the resources to finesse the technology and master system efficiencies in a smarter way.

The remarkable element is that a neutrino source can be fabricated producing extraordinary number of detection events at cosmic distances. The signal to noise ratio at Mev may not even be a problem if the detection rate is high. There may be waste heat from the operation of this technology, even if in the infrared it will look a bit odd.

Acknowledgement: I am grateful to Jason Wright for technical comments and Paul Gilster for a review of the paper.

This paper was inspired by the Technosignatures Workshop in September 2018 [28].

Illustrations by Douglas Potter.

*Email  al_jackson@aajiv.net

# Appendix

Engineering physics problems an advanced civilization has to solve:

(1) The point-like nature of the transmitting aperture can be estimated. In the paper by Matsunaga and Yamamoto [25] , calculate the Einstein angle,

$$\theta_e = \sqrt{\frac{4GMd_{ls}}{c^2 d_s^2}} \quad (13)$$

M is the mass on the lens , $d_{ls}$ is the distance of the transmitter from the lens and $d_l$ is the distance of the lens from the target, then the size of the transmitting aperture (not the size of the transmitter) is

$$s = \frac{d_s \theta_e}{\sqrt{w}} \quad (14).$$

That is approximately 40 microns for the thought experiment here, the realization of which will require the ingenuity of a K2 civilization.

(2) The modeling in section 2 is highly idealized, the wave optics is calculated on the basis of assuming plane waves and that the gravitational lens is spherical. For a non-rotating black hole sphericity is satisfied, however for a neutron star small departures from a sphere are possible. Accretion of matter, crust stresses and magnetic fields can cause a deformation. If the departure from a perfect sphere is measured by a small parameter ε then it can be shown that the lens magnification gain (8) is reduced by (26)

$$gain_{reduced} \approx \frac{.25}{\sqrt{\varepsilon\, gain_0}} \quad , \varepsilon\, gain_0 \gg 1.$$

A deformation of a neutron star will induce a quadrupole moment, the maximum ε can be as large as $10^{-4}$, however the induced gravitational radiation will dissipate the distortion from a sphere on the order of seconds.

(3) Orbital configuration. At 1000 neutron star radii the orbital speed is .01c at about .1 radian/sec which will be a guidance , navigation and control problem, only a civilization with interstellar flight could solve. To be noted, at 100 neutron star radii the tidal forces on a solid vehicle of steel will be quite small , for a black hole the constellation would have to be inside 10 Schwarzschild radii to notice departures from Newtonian mechanics.